\documentclass[a4paper]{article}
\usepackage{amsmath,graphicx,amssymb,delarray,subfigure,verbatim, bbding}
\usepackage{diagbox, makecell, mathtools}
\usepackage{bbding}
\usepackage{diagbox, makecell}
\usepackage{multirow}
\usepackage{mathrsfs}
\usepackage{color}
\usepackage{booktabs}
\usepackage{INTERSPEECH2022}
\usepackage{threeparttable}

\makeatletter
\renewcommand\normalsize{%
	\abovedisplayskip 3\p@ \@plus3\p@ \@minus3\p@
	\abovedisplayshortskip \z@ \@plus3\p@
	\belowdisplayshortskip 6\p@ \@plus3\p@ \@minus3\p@
	\belowdisplayskip \abovedisplayskip
	\let\@listi\@listI}
\makeatother

\title{Optimizing Shoulder to Shoulder: A Coordinated Sub-Band Fusion Model for Real-Time Full-Band Speech Enhancement}
\name{Guochen Yu$^{1, 2}$, Andong Li$^{2}$, Wenzhe Liu$^{2}$, Chengshi Zheng$^{2}$\thanks{Chengshi Zheng is the corresponding author}, Yutian Wang$^{1}$, Hui Wang$^{1}$}
\address{
	$^1$State Key Laboratory of Media Convergence and Communication, Communication University of China, Beijing, China\\
	$^2$Institute of Acoustics, Chinese Academy of Sciences, Beijing, China
}
\email{ \{yuguochen, wangyutian, hwang\}@cuc.edu.cn, \{liandong, liuwenzhe, cszheng\}@mail.ioa.ac.cn}
\begin{document}
\maketitle
\begin{abstract}
Due to the high computational complexity to model more frequency bands, it is still intractable to conduct real-time full-band speech enhancement based on deep neural networks. Recent studies typically utilize the compressed perceptually motivated features with relatively low frequency resolution to filter the full-band spectrum by one-stage networks, leading to limited speech quality improvements. In this paper, we propose a coordinated sub-band fusion network for full-band speech enhancement, which aims to recover the low- (0-8 kHz), middle- (8-16 kHz), and high-band (16-24 kHz) in a step-wise manner. Specifically, a dual-stream network is first pretrained to recover the low-band complex spectrum, and another two sub-networks are designed as the middle- and high-band noise suppressors in the magnitude-only domain. To fully capitalize on the information intercommunication, we employ a sub-band interaction module to provide external knowledge guidance across different frequency bands. Extensive experiments show that the proposed method yields consistent performance advantages over state-of-the-art full-band baselines.

\end{abstract}
\vspace{-0.2cm}
\section{Introduction}
\vspace{-0.2cm}
Speech enhancement (SE) aims to rehabilitate the target speech from noise-corrupted mixtures~{\cite{loizou2013speech}}. Recently, deep neural network (DNN) based SE approaches have shown remarkable performance in suppressing highly non-stationary noise over traditional statistical signal processing based methods, especially under low signal-to-noise ratio (SNR) conditions~{\cite{wang2018supervised}}. However, due to the exorbitant computational cost of stepping toward higher frequency bands, most existing DNN-based SE schemes are restricted to the scenarios of narrow- or wide-band speech signals with a sampling rate of 8000 Hz or 16000 Hz. Note that the unit of frequency bands is abbreviated to `kHz' to avoid concept confusion with the unit of the sampling rate in the remainder of this paper.

Instead of performing on the Fourier spectrum directly, previous works usually adopt coarse-grained psycho-acoustically motivated features as the inputs for full-band (sampled at 48000 Hz) SE~{\cite{valin2018hybrid, valin2020perceptually, schroter2021deepfilternet, ge2022percepnet+}}, and thus reduce the overall system complexity. In~{\cite{valin2018hybrid}}, 22-dimensional Bark-frequency cepstral coefficients (BFCC) defined in the Bark-scale were adopted as input features and 22 ideal critical band gains were mapped as the target. More recently, according to the human hearing equivalent rectangular bandwidth (ERB) scale, PercepNet developed a perceptual band representation with only 34 spectral bands~{\cite{valin2020perceptually}}. Although these approaches can lower the frequency-wise feature dimension and the overall calculation complexity can then be decreased, the frequency resolution of the spectrum in Bark scale and that in ERB scale are much smeared than the original Fourier spectrum, leading to inaccurate spectrum recovery and information loss among frequency bands. Moreover, due to the fact that the wide-band (0-8 kHz) tend to have more energy, tonalities, and harmonics than the higher-frequency bands (8-24 kHz), simultaneously modeling 0-8 kHz and 8-24 kHz frequency bands by a single network may severely degrade SE performance, especially in the high-frequency region~{\cite{zhang2022two}}.

To resolve the aforementioned problems, this paper proposes a coordinated \textbf{S}ub-band \textbf{F}usion network, dubbed \textbf{SF-Net}, for real-time full-band SE using short-time Fourier transform (STFT) features. To be specific, we split the original full-band spectrum into low-band (LB), middle-band (MB) and high-band (HB) spectra, and three sub-networks are elaborately devised to cope with them accordingly. Motivated by the decoupling-mechanism in recent phase-aware wide-band SE methods~{\cite{li2021two, li2021simultaneous, li2022glance, yu2021dual, yu2022dbt}}, we first pre-train a dual-stream network, called \textbf{DSLB-Net}, to address the LB complex spectrum (0-8 kHz) recovery, which mainly comprises of a magnitude estimation network (ME-Net) and a complex purification network (CP-Net). From the complementary perspective, ME-Net aims to coarsely suppress noise components in the magnitude domain, while CP-Net is established to compensate for the missing spectral details and implicitly recover phase information in the complex domain. Then, we integrate the pretrained DSLB-Net with another two higher-band masking networks, namely \textbf{MBM-Net} and \textbf{HBM-Net}, to tackle the 8-16 kHz and 16-24 kHz bands. Due to the fact that speech in higher frequency bands contains lower energies and fewer harmonics, we only map the magnitude gain and retain the phase unaltered for the 8-24 kHz bands. Besides, to capitalize on the implicit correlations among different frequency bands, a sub-band interaction module is devised within the MBM-Net and HBM-Net, which aims to extract the knowledge from the estimated LB spectrum as guidance. Finally, the estimated low-, middle- and high-band spectra are fused to obtain the full-band signal. Comprehensive experiments on two public benchmarks well validate the superiority of the proposed method in various evaluation metrics.

The remainder of the paper is organized as follows. In Section~{\ref{Sec2}}, the proposed framework is described in detail. The experimental setup is presented in Section~{\ref{Sec3}}, while Section~{\ref{Sec4}} gives the results. Some conclusions are drawn in Section~{\ref{Sec5}}.
\begin{figure*}[t]
	\centering
	\centerline{\includegraphics[width=1.85\columnwidth]{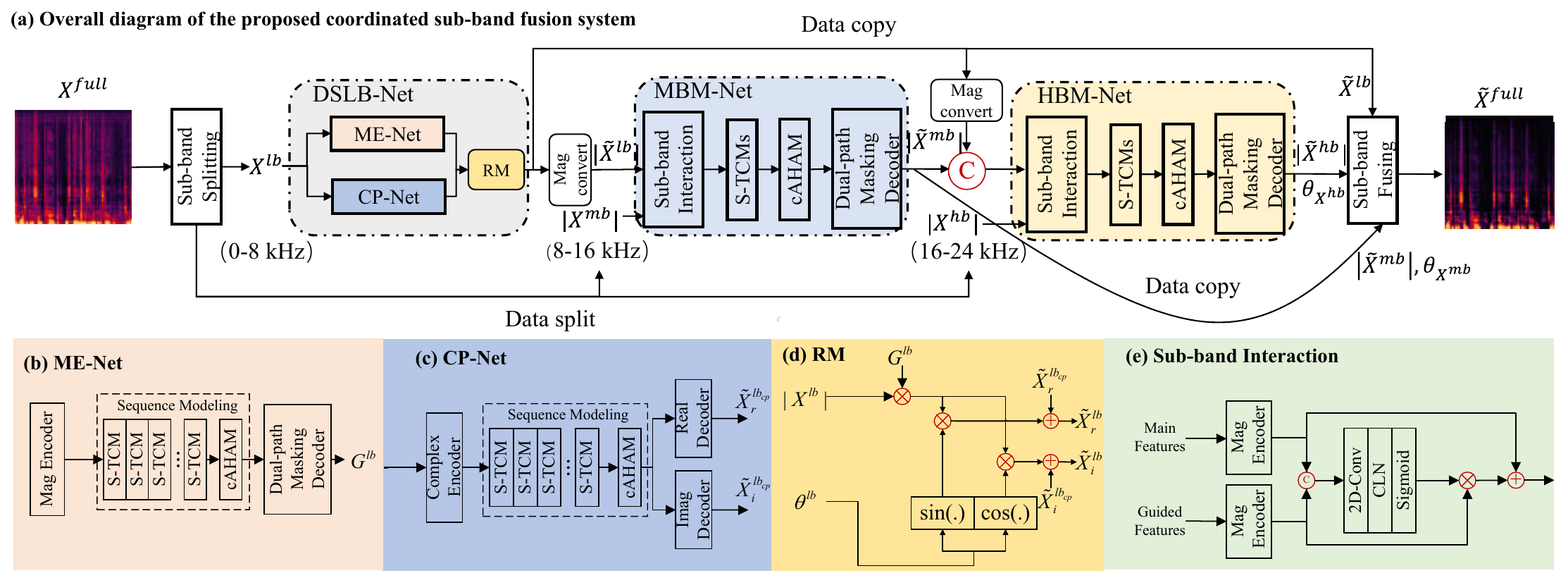}}
	\vspace{-0.2cm}
	\caption{ The overall diagram of the proposed system. Different modules are indicated with different colors for better visualization. \textcolor{red}{$\textcircled{c}$} denotes the concatenation operation in the channel axis. }
	\label{fig:diagram_system}
	\vspace{-0.5cm}
\end{figure*}
\vspace{-0.3cm}
\section{Methodlogy\label{Section2}}
\label{Sec2}
\vspace{-0.15cm}
\subsection{Collaborative dual-stream low-band SE \label{Section23}}
\vspace{-0.2cm}
The overall diagram of the proposed approach is shown in Figure~{\ref{fig:diagram_system}}(a), which is comprised of three sub-networks, namely DSLB-Net, MBM-Net, and HBM-Net. As speech contains more harmonics and semantic information in the frequency range of 0 to 8 kHz, we first employ DSLB-Net to eliminate the noise and recover the clean complex spectrum in the LB regions. Inspired by the efficacy of decoupling-style SE methods~{\cite{li2021two, li2021simultaneous, yu2021dual, yu2022dbt}}, we decouple the original complex spectrum estimation into spectral magnitude and phase optimization, and adopt a dual-stream network to collaboratively estimate the magnitude and residual complex components of the low-band (0-8 kHz) spectrum in parallel, which consists of a magnitude estimation network (ME-Net), a complex spectrum purification network (CP-Net), and a reconstruction module (RM), as illustrated in Figure~{\ref{fig:diagram_system}}(b), (c) and (d), respectively. The input features of ME-Net and CP-Net are denoted as $\lvert{{X}^{lb}}\rvert \in \mathbb{R}^{ T\times F\times 1}$ and $X_{com}^{lb} = Cat({X}^{lb}_{r},{X}^{lb}_{r})\in \mathbb{R}^{T\times F\times 2}$, where $\left\{T, F\right\}$ denote the number of frames and that of frequency bins, respectively.

In ME-Net, given noisy low-band spectral magnitude $\left | X^{lb} \right |$, the network estimates a real-valued gain function $G^{lb}$, which aims at coarsely filtering out the dominant noise. Then the denoised LB spectral magnitude is coupled with the original noisy phase to obtain the coarse-estimated real and imaginary (RI) spectrum. As the supplementation, we leverage CP-Net to purify the spectral structures and also recover the phase information. Rather than explicitly estimating the complex spectrum from scratch, CP-Net is designed for residual mapping, which alleviates the overall burden of this network. Taking the above estimations as the input, we reconstruct the estimated spectrum via the proposed RM. The whole procedure is thus formulated as:
\vspace{-0.0cm}
\begin{gather}
	\label{eqn1}
	\lvert\widetilde{X}^{lb_{me}}\rvert = \lvert{X^{lb}}\rvert \otimes G^{lb},\\
	\widetilde{X}^{lb_{me}}_{r} = \vert\widetilde{X}^{lb_{me}}\rvert \otimes \cos\left( \theta_{X^{lb}}\right),\\
	\widetilde{X}^{lb_{me}}_{i} = \lvert\widetilde{X}^{lb_{me}}\rvert \otimes \sin\left( \theta_{X^{lb}} \right), \\
	\widetilde{X}^{lb}_{r} = \widetilde{X}^{lb_{me}}_{r} + \widetilde{X}^{lb_{cp}}_{r}, \\
	\widetilde{X}^{lb}_{i} = \widetilde{X}^{lb_{me}}_{i} + \widetilde{X}^{lb_{cp}}_{i},
\end{gather}
where $\left\{\widetilde{X}^{lb_{cp}}_{r}, \widetilde{X}^{lb_{cp}}_{i}  \right\}$ denote the output residual RI components of CP-Net and $\left\{\widetilde{X}^{lb}_{r}, \widetilde{X}^{lb}_{i}\right\}$ denote the final merged estimation of clean RI components. $G^{lb}$ and $\theta_{X^{lb}}$ denotes the estimated gain of ME-Net and the LB noisy phase, respectively.
 $\otimes$ is the element-wise multiplication operator. 
\vspace{-0.25cm}
\subsection{Sub-band fusion for full-band SE \label{Section21}}
\vspace{-0.1cm}
Based on the fact that the frequency bands ranging from 8-24 kHz tend to contain less speech information, we further employ two light-weight sub-networks, namely MBM-Net and HBM-NET, as the noise suppressors for the middle-band and high-band, respectively. To reduce the computational burden of the network and evade the implicit compensation effect between magnitude and phase~{\cite{wang2021compensation}, we only consider the magnitude and retain the phase unaltered in these two bands. Besides, we model the correlations among LB, MB, and HB via the proposed interaction module, where the estimated LB features are employed to guide the spectrum recovery of MB and HB.

To be specific, the estimated LB spectral magnitude by DSLB-Net is fed into MBM-Net and HBM-Net along with the noisy MB and HB spectral magnitude. Taking MBM-Net as an example, as shown in Figure~{\ref{fig:diagram_system}}(e), we adopt two encoders to extract the magnitude features from noisy MB spectra and the pre-estimated LB spectra. After that, the features are concatenated together and fed into the mask block to derive the gain function, which aims to automatically learn to filter and preserve different areas of the guided LB feature. Then we sum the extracted FB feature and the filtered version from LB to yield the interacted representation and feed for the latter modules. For HB regions, likewise, the denoised LB and MB spectra are concatenated in the channel axis to obtain the guided input features. With the proposed interaction module, external knowledge from DSLB-Net can effectively propagate to MBM-Net and HBM-Net regions and gradually guide the spectrum recovery.

In summary, the operation stream of the middle- and high-band modeling can be formulated as:
\begin{gather}
	\label{eqn6}
	|\tilde{X}^{mb}| = \lvert{X^{mb}}\rvert \otimes \mathcal{G}^{mb}\left( |X^{mb}|, \lvert\widetilde{X}^{lb}\rvert; \Phi_{1} \right),\\
	|\tilde{X}^{hb}| = \lvert{X^{hb}}\rvert \otimes \mathcal{G}^{hb}\left( |X^{hb}|,\lvert\widetilde{X}^{lb}\rvert, \lvert\widetilde{X}^{mb}\rvert; \Phi_{2} \right),\\
	\tilde{X}^{mb} = |\tilde{X}^{mb}| \exp(j\theta_{X^{mb}}), \tilde{X}^{hb} = |\tilde{X}^{hb}| \exp(j\theta_{X^{hb}}),
\end{gather} 	
where $|\tilde{X}^{mb}|$ and $|\tilde{X}^{hb}|$ denote the denoised MB and HB outputs, respectively. $\mathcal{G}^{mb}$ and $\mathcal{G}^{hb}$ denote the mapping functions of MBM-Net and HBM-Net with parameter set $\Phi_{\left(\cdot\right)}$. $\theta_{X^{mb}}$ and $\theta_{X^{hb}}$ denote the noisy MB and HB phase, respectively. Finally, we stack the estimated three sub-band estimation along the frequency axis to obtain the full-band spectrum. Note that we average the overlapped bands in the MB and HB regions.
\vspace{-0.3cm}
\subsection{Network architecture \label{Section22}}
\vspace{-0.15cm}
The detailed architectures of DSLB-Net, MBM-Net, and HBM-Net are shown in Figure~{\ref{fig:diagram_system}}. Similar to~{\cite{li2021two, li2021simultaneous}}, we employ a classical convolutional encoder-decoder topology~{\cite{zhao2020noisy} for these sub-networks, where multiple temporal convolution modules (TCMs) and a causal adaptive hierarchical attention module (cAHAM)~{\cite{yu2021dual}} are stacked in the bottleneck for sequence modeling. More specifically, the detailed structures of ME-Net and CP-Net in DSLB-Net are shown in Figure~{\ref{fig:diagram_system}}(b) and (c), in which ME-Net utilizes a magnitude encoder and a dual-path masking decoder and CP-Net utilizes a complex encoder and two decoders to recover both RI components.
	
Taking ME-Net as an example, the encoder consists of five downsampling blocks, each of which consists of a convolutional layer, a normalization layer, and PReLU, with kernel size being (2, 3) in the time and frequency axes except (2, 5) in the first block. The number of channels remains 64 by default, and the stride is set to (1, 2) to gradually halve the frequency size. To enable streaming inference, the features are normalized with cumulative layer norm (cLN)~{\cite{luo2019conv}}, where the statistics are adaptively updated in a frame-wise manner. The dual-path masking decoder consists of five symmetrical deconvolutional layers and a dual-path mask module, which is performed to obtain the magnitude spectral gain by a 2-D convolution and a dual-path tanh/sigmoid nonlinearity operation similar to~{\cite{yu2021dual}. 

Inspired by~{\cite{pandey2019tcnn}, four groups of squeezed TCMs (S-TCMs) are employed for sequence modeling, each of which stacks six S-TCMs with increasing dilation rates $d$ to obtain a large temporal receptive field, \emph{i.e.}, $d = \{1, 2, 4, 8, 16, 32\}$. To reduce the computational burden, the parameter weights of S-TCMs are shared in ME-Net and CP-Net. Similar to our previous study~{\cite{yu2021dual, yu2021cyclegan}, we utilize a casual AHAM~{\cite{yu2021dual}} to integrate all intermediate features and global hierarchical contextual information during sequence modeling, where the average pooling layer is adopted along the frequency axis and retain the time resolution unaltered to satisfy the real-time requirement. For MBM-Net and HBM-Net, we leverage a similar structure as ME-Net, which aims at filtering out the dominant noise in MB and HB regions, respectively. 
\vspace{-0.35cm}
\subsection{Loss function \label{Section24}}
\vspace{-0.15cm}
In our SF-Net, we adopt a two-stage training pipeline to recover low-band and full-band spectra progressively. First, we pretrain DSLB-Net with MSE until convergence. The loss function involves both RI and magnitude estimation, given as:
\begin{gather}		
	\mathcal{L}_{lb}^{RI}=\left \|\widetilde{X}^{lb}_r-S^{lb}_r \right \|_{F}^2 +\left \|\widetilde{X}^{lb}_i-S^{lb}_i \right \|_{F}^2,\\
	\mathcal{L}_{lb}^{Mag}=\left \| \lvert \tilde X^{lb} \rvert - \lvert S^{lb} \rvert \right \|^{2}_{F},\\
	\mathcal{L}_{lb}=\mu \mathcal{L}_{lb}^{RI}+(1-\mu ) \mathcal{L}_{lb}^{Mag} ,
\end{gather}
where $\mathcal{L}_{lb}^{Mag}$ and $\mathcal{L}_{lb}^{RI}$ denote the loss terms toward magnitude and RI components, respectively. $\lvert S^{lb}\rvert$ denotes the target LB spectral magnitude. $S^{lb}_r$ and $S^{lb}_i$ represent the RI components of target LB regions. With the internal trial, we empirically set $\mu= 0.5$ in the following experiments.

In the second stage, we couple the pretained DSLB-Net with MBM-Net and HBM-Net, and train them jointly. The overall loss can be given by:
\begin{gather}
	\mathcal{L}_{full}=\alpha \mathcal{L}_{lb} + \mathcal{L}_{mb}^{Mag} + \mathcal{L}_{hb}^{Mag}
\end{gather}
where $\mathcal{L}_{mb}^{Mag}$ and $\mathcal{L}_{hb}^{Mag}$ denote the loss functions for MF-Net and HF-Net in the magnitude domain, while $\mathcal{L}_{full}$ represents the full loss function of the second stage.  We empirically find that $\alpha= 0.1$ suffices in our evaluation.
\vspace{-0.3cm}
\section{Experimental setup\label{Section3}}
\label{Sec3}
\vspace{-0.3cm}
\subsection{Datasets\label{Section31}}
\vspace{-0.1cm}
To verify the effectiveness of the proposed model, we conduct extensive experiments on two public full-band benchmarks, namely VoiceBank + DEMAND dataset~\cite{valentini2016investigating} and the ICASSP 2022 DNS-Challenge (DNS-2022) dataset~\cite{dubeyicassp}, where all the utterances are sampled at 48000 Hz. 

\textbf{VoiceBank + DEMAND}: The dataset is a selection of the VoiceBank corpus~{\cite{veaux2013voice}} with 28 speakers for training and another 2 unseen speakers for testing. The training set consists of 11,572 mono audio samples, while the test set contains 824 utterances by 2 speakers (one male and one female). For the training set, all the audio samples are mixed with one of the 10 noise types, including two artificial noise processes and eight real noise recordings taken from the Demand database~{\cite{thiemann2013diverse}}.

\textbf{DNS-Chanllenge}: We further train and evaluate our model on the DNS-2022 dataset{\footnote{https://github.com/microsoft/DNS-Challenge}}, which consists of various clean speech, noise clips, and room impulse responses (RIRs) to simulate practical acoustic scenarios. For this dataset, we totally generate around 600 hours of noisy-clean pairs. To generate reverberant-noisy training data, we use 248 real and 60,000 synthetic RIRs from openSLR26 and openSLR28 datasets~{\cite{ko2017study}}. As only late reverberation degrades the speech quality/intelligibility~{\cite{zhao2020monaural}}, we preserve both anechoic and early reverberation components as the training target. During each mixing process, the clean speech is convolved with a randomly selected RIR, and is then mixed with the noise in the SNR range of $\left(-5\rm{dB}, 15\rm{dB}\right)$. We evaluate the model performance upon the DNS-2022 blind test set, which includes 859 real test clips.
\vspace{-0.35cm}
\subsection{Implementation setup\label{Section32}}
\vspace{-0.15cm}
The 20ms Hanning window is utilized, with 50\% overlap between adjacent frames. To extract the features, 960-point FFT is utilized and 481-dimension spectral features are obtained. Due to the efficacy of power compression in both dereverberation and denoising tasks~{\cite{li2021importance}}, we conduct the power compression toward the spectral magnitude while remaining the phase unaltered, and the compression factor is set to $\beta=0.5$. All the models are optimized using Adam ($\beta_{1}=0.9$, $\beta_{2}=0.999$)~{\cite{kingma2014adam}}. In the first stage, the initialized learning rate (LR) is set to 1e-3 for DSLB-Net. In the
second stage, DSLB-Net is fine-tuned with LR of 1e-4, while LR is set to 1e-3 for MBM-Net and HBM-Net. The batch size is set to 16 at the utterance level. \textbf{The processed samples are available online, where the source code will be released soon.}{\footnote{https://github.com/yuguochencuc/SF-Net}
\vspace{-0.35cm}
\section{Experimental results and discussion\label{Section4}}
\label{Sec4}
\vspace{-0.15cm}		
In this study, we mainly adopt wide-band PESQ~{\cite{rix2001perceptual}}, STOI~{\cite{taal2010short}}, CSIG, CBAK, and COVL~{\cite{hu2007evaluation}} to evaluate low-band SE performance, while the segmental SNR (SSNR) and SDR~{\cite{vincent2007first}} are employed for full-band SE evaluation. Higher values indicate better performance. 
\renewcommand\arraystretch{1.0}
\begin{table}[t!]
	\caption{Ablation study $\emph{w.r.t.}$ dual-stream structure, sub-band fusion strategy, cAHAM and sub-band interaction module.}
	\setlength\tabcolsep{4pt}
	\centering
	\Large
	\vspace{-0.3cm}
	\begin{threeparttable}
		\resizebox{\columnwidth}{!}{
			\begin{tabular}{c|ccc|cc|cc}
				\toprule
				\multirow{2}*{Models} &\multirow{2}*{Feat.} &{Para.} &{MACs} & \multicolumn{2}{c|}{\textbf{wide-band$^*$}}& \multicolumn{2}{c}{\textbf{full-band}}\\
				\cline{5-6} \cline{7-8}
				&  &(M) &(G/s) &{PESQ$\uparrow$}   &{STOI(\%)$\uparrow$}    &SDR(dB)$\uparrow$ &SSNR(dB)$\uparrow$\\
				\hline
				Noisy &\makecell[c]{--} & \makecell[c]{--} & \makecell[c]{--} &1.97 &92.1 &8.42 & 1.71 \\ 
				\hline
				\multicolumn{8}{c}{\textbf{One-stage full-band approaches}} \\ 		
				\hline
				ME-Net (full)  & Mag & 2.18 & 3.14 &2.72 &94.0 &16.26 &7.23  \\ 
				CP-Net (full)  & RI &2.89 &5.57  &2.67 &93.1 &15.38& 7.34 \\			
				DS-Net (full)  & Mag+RI &3.30 &8.71 & 2.78 & 94.3 & 18.86 &9.12 \\  			
				\hline
				\multicolumn{8}{c}{\textbf{Sub-band fusion approaches}} \\ 	
				\hline
				ME-SF-Net  & Mag &6.42 &3.73 &2.83 &93.6 & 16.74 &8.13  \\ 
				CP-SF-Net  &RI &7.22 &6.04 &2.90 &94.2 &17.04 &8.28 \\
				SF-Net (\textbf{Pro.})     & RI+Mag & 6.98 & 5.62 & \textbf{3.02} & \textbf{94.5} & \textbf{19.90} & \textbf{9.69} \\  
				~ - cAHAM    & RI+Mag & 6.82 & 5.49  & 2.94 & 94.2 & 19.67 &9.13 \\  
				~~ - Inter.   &RI+Mag & 6.49  &5.11  & 2.93  &94.4 &18.99 &8.80  \\
				\bottomrule
			\end{tabular}
		}
		\begin{tablenotes}
			\footnotesize
			\item *: Calculated on downsampled speech at 16000 Hz.  
		\end{tablenotes}	
	\end{threeparttable}
	\label{tbl:ablation_study}
	\vspace{-0.8cm}	
\end{table}
\vspace{-0.35cm} 
\subsection{Ablation study \label{Section41}}
\vspace{-0.15cm} 
We first conduct ablation studies to investigate the effects of the proposed sub-band fusion strategy, dual-stream structure, cAHAM and the sub-band interaction module, including a) three one-stage full-band SE approaches, \emph{i.e.}, the magnitude estimation network (dubbed ME-Net (full)), the complex purification network (dubbed CP-Net (full)), and the dual-stream network (dubbed DS-Net (full)), b) three sub-band fusing-based methods, namely ME-SF-Net, CP-SF-Net and the proposed coordinated sub-band fusion model (SF-Net), c) SF-Net without the causal adaptive hierarchical attention module (-cAHAM) and sub-band interaction module (- Inter.). Due to the lack of PESQ evaluation for full-band speech, we downsample the outputs of all models at 16000 Hz and measure wide-band PESQ and STOI, while SSNR and SDR are measured on the full-band SE.

Quantitative results are presented in Table~{\ref{tbl:ablation_study}}, one can have the following observations. First, among the one-stage full-band approaches, DS-Net dramatically outperforms other single-stream approaches in terms of wide-band and full-band SE performance. For example, DS-Net provides average 0.21, 1.2\%, 3.38\rm{dB} and 1.78\rm{dB} improvements than CP-Net in PESQ, STOI, SDR and SSNR, respectively. This reveals the effectiveness of the proposed dual-stream SE topology in improving speech quality. Second, compared with the one-stage full-band SE, when employing the sub-band fusion strategy, considerable performance improvements in terms of both wide-band and full-band cases can be obtained. Besides, we also provide the model size and the number of multiply-accumulate operations (MACs) per second, as shown in Table~{\ref{tbl:ablation_study}}. Note that although SF-Net suffers more parameters than DS-Net (full), it achieves fewer MACs and remarkably better performance. This indicates the effectiveness of the proposed sub-band splitting and fusion strategy. Finally, without using cAHAM and sub-band interaction module, consistent performance degradations in both wide-band and full-band speech are observed, which emphasize the significance of cAHAM and sub-band interaction in improving speech quality.

Moreover, spectrograms of noisy utterance, clean utterance and enhanced utterances by DS-Net and SF-Net are presented in Figure~{\ref{fig:visualization}} (a)-(d). Focusing on the red and green boxes in Figure~{\ref{fig:visualization}} (b) and (c), one can see that SF-Net can better suppress background noise than DS-Net in the high-frequency regions, while more spectral details can be restored in the low-frequency regions. This indicates the superiority of SF-Net in improving speech quality in both low- and high-band cases.

\begin{figure}[t]
	\centering
	\centerline{\includegraphics[width=1.1\columnwidth]{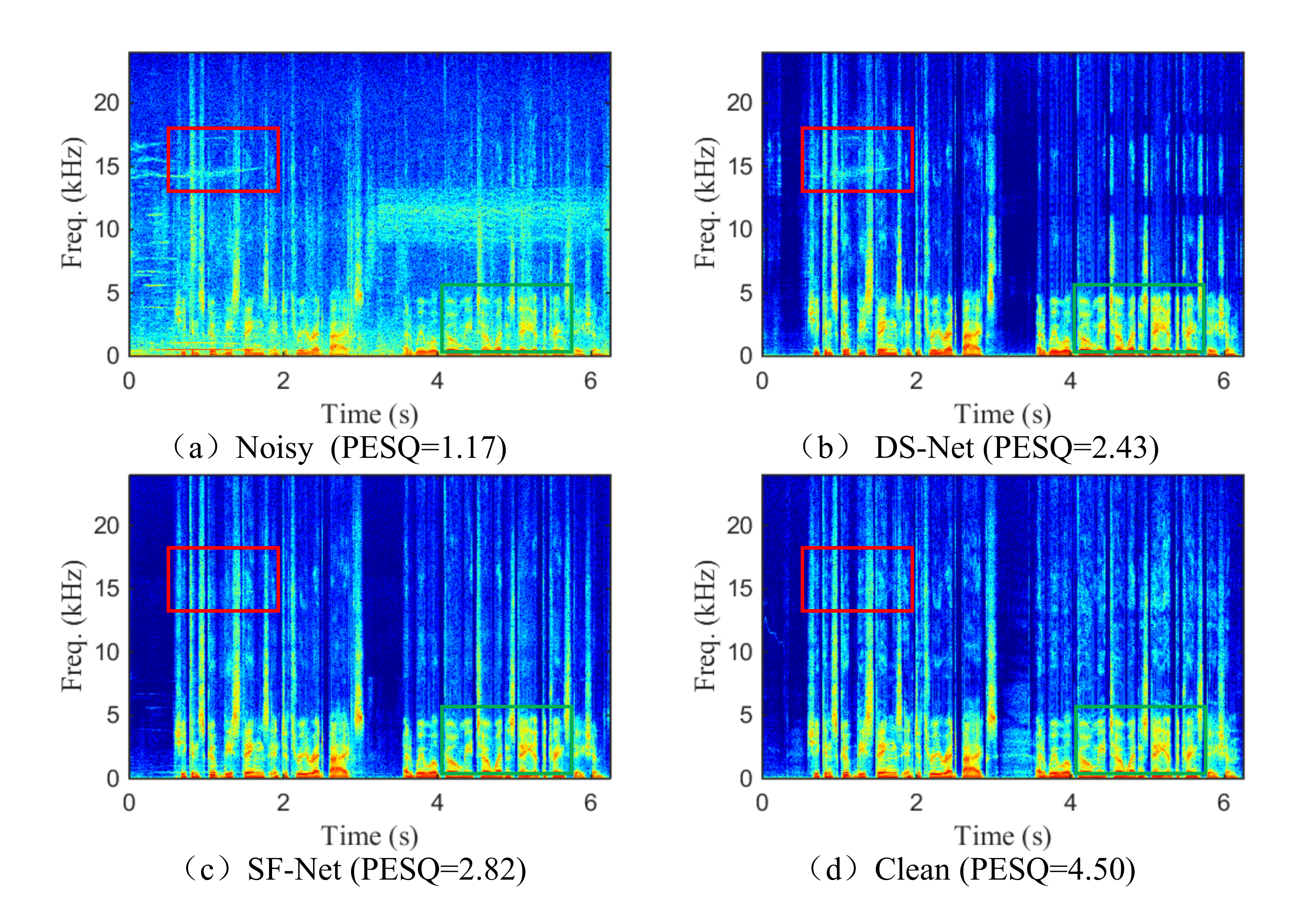}}
	\vspace{-0.5cm}	
	\caption{Visualization of spectrograms. (a) Noisy utterance (b) Enhanced utterance by DS-Net. (c) Enhanced utterance by SF-Net. (d) Clean utterance. }
	\label{fig:visualization}
	\vspace{-0.8cm}		
\end{figure}

\vspace{-0.3cm}
\subsection{Comparison with full-band SOTA methods\label{Section42}}
\vspace{-1mm}
The best configuration of SF-Net in Table~{\ref{tbl:ablation_study}} is chosen to compare with other SOTA baselines, whose results are presented in Table~{\ref{tab:vctk}}. For VoiceBank + DEMAND dataset, six full-band and one super-wideband SE approaches are selected as the baselines, namely GCRN (full-band version)~{\cite{tan2019learning}}, RNNoise~{\cite{valin2018hybrid}}, PercepNet~{\cite{valin2020perceptually}}, CTS-Net (full-band version)~{\cite{li2021two}}, DeepFilterNet~{\cite{schroter2021deepfilternet}}, DMF-Net~{\cite{yu2022dmf}} and S-DCCRN (super-wide band)~{\cite{lv2021s}}. Note that we re-implement GCRN and CTS-Net (full) with power compression for a fair comparison, while we directly use the reported results of other baselines. From Table~{\ref{tab:vctk}}, several observations can be made. First, compared with the compressed psycho-acoustically feature-based methods, SF-Net provides considerably better performance in all objective metrics. Second, the proposed system dramatically outperforms the previous one-stage full-band baselines, demonstrating the effectiveness of the proposed sub-band fusion strategy. Third, compared with our preliminary decoupling-style multi bands fusion model (\emph{i.e.}, DMF-Net), SF-Net achieves further improvements with fewer trainable parameters, especially in the background noise suppression (CBAK) and overall quality (COVL). This indicates that the utilization of the proposed dual-stream network can realize better speech recovery and noise suppression.

We further conduct evaluations on the DNS-2022 blind test set, whose results are presented in Table~{\ref{tab:dnsmos}}. Due to the lack of clean speech as the reference, we utilize the non-intrusive subjective evaluation metrics to evaluate the subjective speech performance, namely DNSMOS P.808~{\cite{reddy2021dnsmos}} and P.835~{\cite{reddy2022dnsmos}}, which are based on ITU-T
P.808~{\cite{naderi20_interspeech}} and P.835~{\cite{naderi21_interspeech}}, respectively. Compared with NSNet2, a standard baseline system for DNS-2022~{\cite{braun2020data}}, the proposed approach yields consistently better performance in speech distortion (SIG), background noise (BAK) and overall quality (OVRL). Besides, SF-Net outperforms DMF-Net in terms of speech distortion and overall quality, while a similar BAK DNSMOS score is observed. This further verifies the superiority of our approach in recovering the speech components in practical acoustic scenarios. 
\renewcommand\arraystretch{1}
\begin{table}[t!]
	\caption{Comparison on VoiceBank + DEMAND dataset. "$-$" denotes that the result is not provided in the original paper.}
	\centering
	\large
	\setlength\tabcolsep{2pt}
	\vspace{-0.3cm}
	\resizebox{\columnwidth}{!}{
		\begin{tabular}{l|c|c|ccccc}
			\toprule
			Models   &Year & Para.(M) &PESQ$\uparrow$ & STOI(\%)$\uparrow$ &CSIG$\uparrow$ &CBAK$\uparrow$ &COVL$\uparrow$ \\
			\hline
			Noisy       &  \makecell[c]{--}  & \makecell[c]{--}  & 1.97 & 92.1 & 3.35 & 2.44 & 2.63   \\
			GCRN (full)~{\cite{tan2019learning}}        & 2019   & 10.59  & 2.71 & 93.8 & 4.12 & 3.23  & 3.41      \\
			RNNoise~{\cite{valin2018hybrid}}      & 2020  & 0.06  & 2.34 & 92.2 & 3.40 & 2.51 & 2.84   \\
			PercepNet~{\cite{valin2020perceptually}}    & 2020     & 8.00  & 2.73  & \makecell[c]{--} & \makecell[c]{--}    & \makecell[c]{--} &\makecell[c]{--}  \\
			CTS-Net (full)~{\cite{li2021two}}        & 2020   & 7.09  & 2.92  & 94.3 & 4.22 & 3.43 & 3.62      \\
			DeepFilterNet~{\cite{schroter2021deepfilternet}} & 2021  & 1.80  & 2.81  & \makecell[c]{--} & \makecell[c]{--}    & \makecell[c]{--} &\makecell[c]{--} \\
			{S}-DCCRN~{\cite{lv2021s}}      & 2022  & 2.34  & 2.84  & 94.0 & 4.03 & 2.97 & 3.43 \\
			DMF-Net~{\cite{yu2022dmf}}     & 2022  & 7.84  & 2.97 & 94.4  & 4.26 & 3.25 & 3.48 \\
			DS-Net (full)    & 2022  & 3.30  & 2.78  & 94.3  & 4.20 & 3.34 & 3.48 \\
			SF-Net (\textbf{Pro.})      & 2022  & 6.98 & \textbf{3.02} & \textbf{94.5}  & \textbf{4.36} & \textbf{3.54} & \textbf{3.67} \\
			\bottomrule
	\end{tabular}}
	\vspace{-0.3cm}		
	\label{tab:vctk}
\end{table}
\renewcommand\arraystretch{0.95}
\begin{table}[t!]
	\centering
	\footnotesize
	\caption{DNSMOS P.835 and P.808 results on DNS-2022 blind test set.}
	\vspace{-0.3cm}
	\scalebox{1}{
		\begin{threeparttable}
			
			\begin{tabular}{lcccc}
				\toprule
				\multirow{2}*{Model}   &\multicolumn{3}{c}{DNSMOS P.835$^*$} &{DNSMOS$^*$}\\
				\cline{2-4}
				& SIG& BAK &OVRL &P.808\\
				
				\midrule
				Noisy     & 4.14 & 2.94 & 3.27 &3.03   \\
				NSNet2~{\cite{braun2020data}}  & 3.87 & 4.21 & 3.58 &3.57 \\
				DMF-Net~{\cite{yu2022dmf}}  & 3.92 & \textbf{4.57} & 3.72 &3.61 \\
				DS-Net (full)  & 4.07 & 4.29& 3.77 &3.65\\
				SF-Net(\textbf{Pro.}) & \textbf{4.15} & 4.49 &\textbf{3.94} &\textbf{3.73} \\
				\bottomrule
			\end{tabular}
			\begin{tablenotes}
				\item *: Calculated on downsampled speech at 16000 Hz.  
			\end{tablenotes}
	\end{threeparttable}}
	\vspace{-0.8cm}
	\label{tab:dnsmos}
\end{table}

\vspace{-0.3cm}
\section{Conclusions\label{Section5}}
\label{Sec5}
\vspace{-0.1cm}

In this paper, we propose a collaborative sub-band fusion approach, dubbed SF-Net, for real-time speech enhancement running on 48 kHz-sampled speech signals. Motivated by the curriculum learning concept, we split the full-band target into three frequency sub-bands, i.e., low-band (0-8 kHz), middle-band (8-16 kHz), and high-band (16-24 kHz), and three chain sub-networks are elaborately designed to recover the full-band clean spectrum by a stage-wise manner. Specifically, conducting on the STFT domain, a dual-stream decoupling-style network is employed to denoise the low-band complex spectrum, and two magnitude-masking based networks are employed to recover the middle- and high-band spectral magnitude. Experimental results demonstrate that the proposed method yields state-of-the-art performance over previous baselines by a large margin.

\vfill\pagebreak
\bibliographystyle{IEEEtran}
\bibliography{myrefs}

\end{document}